\documentclass[10pt,twoside,a4paper]{article}
 
\usepackage{amsmath, amsthm, amsfonts}

\usepackage{footnote,url}

\usepackage{graphicx}

\usepackage[T1]{fontenc}
\usepackage{lmodern}

\usepackage{float} 

\usepackage{authblk}

\usepackage[numbers]{natbib} 

\usepackage{geometry}
\newcommand{\hashtag}[1]{\textsl{\##1}}

\begin{document}
\title{\huge \hashtag{lockdown}: network-enhanced emotional profiling at the times of COVID-19}

\author[1,$\star$]{Massimo Stella}
\author[2]{Valerio Restocchi}
\author[3]{Simon De Deyne}

\affil[1]{\small Complex Science Consulting, 73100 Lecce, Italy}
\affil[2]{School of Informatics, University of Edinburgh, EH8 9AB Edinburgh, Scotland, UK}
\affil[3]{Human Complex Data Hub, School of Psychological Sciences, University of Melbourne, 3010 Victoria, Australia}
\affil[$\star$]{Corresponding author: massimo.stella@inbox.com}

\maketitle

\abstract{
\small
The COVID-19 pandemic forced countries all over the world to take unprecedented measures like nationwide lockdowns. To adequately understand the emotional and social repercussions, a large-scale reconstruction of how people perceived these unexpected events is necessary but currently missing. We address this gap through social media by introducing MERCURIAL (Multi-layer Co-occurrence Networks for Emotional Profiling), a framework which exploits linguistic networks of words and hashtags to reconstruct social discourse describing real-world events. We use MERCURIAL to analyse 101,767 tweets from Italy, the first country to react to the COVID-19 threat with a nationwide lockdown. The data were collected between 11th and 17th March, immediately after the announcement of the Italian lockdown and the WHO declaring COVID-19 a pandemic. Our analysis provides unique insights into the psychological burden of this crisis, focussing on: (i) the Italian official campaign for self-quarantine (\hashtag{iorestoacasa}), (ii) national lockdown (\hashtag{italylockdown}), and (iii) social denounce (\hashtag{sciacalli}). Our exploration unveils evidence for the emergence of complex emotional profiles, where anger and fear (towards political debates and socio-economic repercussions) coexisted with trust, solidarity, and hope (related to the institutions and local communities). We discuss our findings in relation to mental well-being issues and coping mechanisms, like instigation to violence, grieving, and solidarity. We argue that our framework represents an innovative thermometer of emotional status, a powerful tool for policy makers to quickly gauge feelings in massive audiences and devise appropriate responses based on cognitive data.\\}



\section{Introduction}
\normalsize

The stunningly quick spread of the COVID-19 pandemic catalysed the attention of worldwide audiences, overwhelming individuals with a deluge of often contrasting content about the severity of the disease, the uncertainty of its transmission mechanisms, and the asperity of the measures taken by most countries to fight it \cite{zarocostas2020fight,cinelli2020covid,gallotti2020assessing,pulido2020twitter}. Although these policies have been seen as necessary, they had a tremendous impact on the mental well-being of large populations \cite{Wang2020psychological} for a number of reasons. Due to lockdowns, many are facing financial uncertainty, having lost or being on the verge of losing their source of income. Moreover, there is much concern about the disease itself, and most people fear for their own health and that of their loved ones \cite{WHO2020mentalhealth}, further fueled by \textit{infodemics} \cite{cinelli2020covid,gallotti2020assessing,zarocostas2020fight}. Finally, additional distress is caused by the inability of maintaining a normal life \cite{Zhu2020mentalhealth}. The extent of the impact of these factors is such that, in countries greatly struck by COVID-19 such as China, the population started to develop symptoms of post-traumatic stress disorder \cite{Wang2020ptsd}.

During this time more than ever, people have shared their emotions on social media. These platforms provide an excellent emotional thermometer of the population, and have been widely explored in previous studies investigating how online social dynamics promote or hamper content diffusion \cite{cinelli2020covid,ferrara2015quantifying,zarocostas2020fight,davis2020phase,ciulla2012beating} and the adoption of specific positive/negative attitudes and behaviours \cite{ferrara2015quantifying,stella2018bots,bail2018exposure}.

\subsection{Research aim}

Building on the above evidence, our goal is to draw a comprehensive quantitative picture of people's emotional profiles, emerging during the COVID-19 crisis, through a cognitive analysis of online social discourse. We achieve this by introducing MERCURIAL (Multi-layer Co-occurrence Networks for Emotional Profiling), a framework that combines cognitive network science \cite{siew2019cognitive,stella2020text,amancio2015probing} with computational social sciences \cite{ferrara2015quantifying,stella2018bots,cinelli2020covid,quercia2011our,mohammad2010emotions}. 
Before outlining the methods and main contributions of our approach, we briefly review existing research on understanding emotions in social media.

\subsection{Past approaches bridging cognitive, computer and network science}

Much of the research on emotions in social media has been consolidated into two themes. On the one hand, there is the data science approach, which mostly focused over large-scale positive/negative sentiment detection \cite{ferrara2015quantifying} and recently identified the relevance of tracing more complex affect patterns for understanding social dynamics \cite{mohammad2018semeval,kleinberg2020measuring,brito2020complex,amancio2015probing}.
On the other hand, cognitive science research makes use of small-scale analysis tools, but explores the observed phenomena in much more detail in the light of its theoretical foundations \cite{plutchik1991emotions,ekman1994nature,hatfield1993emotional}. Specifically, in cognitive science the massive spread of semantic and emotional information through verbal communication represent long-studied phenomena, known as \textit{cognitive contagion} \cite{hatfield1993emotional} and \textit{emotional contagion} \cite{hatfield1993emotional,barsade2002ripple,ekman1994nature}, respectively. This research suggests that ideas are composed of a cognitive component and an emotional content, much alike viruses containing the genomic information necessary for their replication \cite{zarocostas2020fight}. Both these types of contagion happen when an individual is affected in their behaviour by an idea.  Emotions elicited by ideas can influence users' behaviour without their awareness, resulting in the emergence of specific behavioural patterns such as implicit biases \cite{ekman1994nature}. Unlike pathogen transmission, no direct contact is necessary for cognitive and emotional contagion to take place, since both are driven by information processing and diffusion, like it happens through social media \cite{kramer2014experimental,frey2019rippling}. In particular, during large-scale events, ripples of emotions can rapidly spread across information systems \cite{frey2019rippling} and have dramatic effects, as it has recently been demonstrated in elections and social movements \cite{jasper2011emotions,stella2018bots,barsade2002ripple}. 
 
At the intersection of data- and cognitive science is emotional profiling, a set of techniques  which enables the reconstruction of how concepts are emotionally perceived and assembled in user-generated content \cite{quercia2011our,mohammad2010emotions,ferrara2015quantifying,mohammad2018semeval,stella2020text,stella2020forma}. Emotional profiling conveys information about basic affective dimensions such how positive/negative or how arousing a message is, and also includes the analysis of more fine-grained emotions such as \textit{fear} or \textit{trust} that might be associated with the lockdown and people's hopes for the future \cite{ferrara2015quantifying,posner2005circumplex,plutchik1991emotions}. 

Recently, an emerging important line of research has shown that reconstructing the knowledge embedded in messages through social and information network models \cite{de2016large,siew2019cognitive,stella2019innovation} successfully highlight important phenomena in a number of contexts, ranging from the diffusion of hate speech during massive voting events \cite{stella2018bots} to reconstructing personality traits from social media \cite{quercia2011our}. Importantly, to reconstruct knowledge embedded in tweets, recent work has successfully merged data science and cognitive science, introducing linguistic networks of co-occurrence relationships between words in sentences \cite{amancio2015probing,mehler2020topic,brito2020complex} and between hashtags in tweets \cite{stella2018bots}. However, an important shortfall of these works is that these two types of networked knowledge representations were not merged together, thus missing on the important information revealed by studying their interdependence. 

\subsection{Main contributions}

We identify three important contributions that distinguish our paper from previous literature, and make a further step towards consolidating \textit{cognitive network science} \cite{siew2019cognitive} as a paradigm suitable to analyse people's emotions. 

First, we introduce a new framework exploiting the interdependence between hashtags and words, addressing the gap previously discussed. This framework, multi-layer co-occurrence networks for emotional profiling (MERCURIAL), combines both the semantic structure encoded through the co-occurrence of hashtags and the textual message to construct a multi-layer lexical network \cite{stella2018distance}. This multi-layer network structure allows us to contextualise hashtags and, therefore, improve the analysis of their meaning. Importantly, these networks can be used to identify which concepts or words contribute to different emotions and how central they are. 

Second, in contrast to previous work, which largely revolved around English tweets \cite{pulido2020twitter,kleinberg2020measuring}, the current study focusses on Italian Twitter messages. There are several reasons why the emotional response of Italians is particularly interesting. Specifically, i) Italy was the first Western country to experience a vast number of COVID-19 clusters;  ii) the Italian government was the first to declare a national lockdown \footnote{\url{https:/en.wikipedia.org/wiki/2020\_Italy\_coronavirus\_lockdown}, Last Access: 20/04/2020}; iii), the Italian lockdown was announced on 10th March, one day before the World Health Organization (WHO) declared the pandemic status of COVID-19. This enables us to address the urgent need of measuring the emotional perceptions and reactions to social distancing, lockdown, and, more generally, the COVID-19 pandemic. 

Third, thanks to MERCURIAL, we obtain richer and more complex emotional profiles that we analyse through the lens of established psychological theories of emotion. This is a fundamental step in going beyond positive/neutral/negative sentiment and to provide accurate insights on the mental well-being of a population. 
To this end, we take into account three of the most trending hashtags, \hashtag{iorestoacasa} (English: "I stay at home"), \hashtag{sciacalli} (English: "jackals"), and \hashtag{italylockdown}, as representative of positive, negative, and neutral social discourse, respectively. We use these hashtags as a starting point to build multi-layer networks of word and hashtag co-occurrence, from which we derive our profiles. Our results depict a complex map of emotions, suggesting that there is co-existence and polarisation of conflicting emotional states, importantly fear and trust towards the lockdown and social distancing. The combination of these emotions, further explored through semantic network analysis, indicates mournful submission and acceptance towards the lockdown, perceived as a measure for preventing contagion but with negative implications over economy. As further evidence of the complexity of the emotional response to the crisis, we also find strong signals of hope and social bonding, mainly in relation to social flash mobs, and interpreted here as psychological responses to deal with the distress caused by the threat of the pandemic.

\subsection{Manuscript outline}

The paper is organised as follows. In the Methods section we describe the data we used to perform our analysis, and describe MERCURIAL in detail. In the Results section we present the emotional profiles obtained from our data, which are then discussed in more detail in the section Discussion. Finally, the last section highlights the psychological implications of our exploratory investigation and its potential for follow-up monitoring of COVID-19 perceptions in synergy with other datasets/approaches.

We argue that our findings represent an important first step towards monitoring both mental well-being and emotional responses in real time, offering policy-makers a framework to make timely data-informed decisions.

\section{Methods}

In this section we describe the methodology employed to collect our data and perform the emotional profiling analysis. First, we describe the dataset and how it was retrieved. Then, we introduce co-occurrence networks, and specifically our novel method that combines hasthag co-occurrence with word co-occurrence on multi-layer networks. Finally, we describe the cognitive science framework we used to perform the emotional profiling analysis on the so-obtained networks. 

\subsection{Data}
We gathered 101,767 tweets in Italian to monitor how online users perceived the COVID-19 pandemic and its repercussions in Italy. These tweets were gathered by crawling messages containing three trending hashtags of relevance for the COVID-19 outbreak in Italy and expressing three different sentiment polarities:
\begin{itemize}
    \item \hashtag{iorestoacasa} (English: "I stay at home"), a positive-sentiment hashtag introduced by the Italian Government in order to promote a responsible attitude during the lockdown;
    \item \hashtag{sciacalli}, (English: "jackals"), a negative sentiment hashtag used by online users in order to address unfair behaviour rising during the health emergency;
    \item \hashtag{italylockdown}, a neutral sentiment hashtag indicating the application of lockdown measures all over Italy.
\end{itemize}{}

We refer to \hashtag{iorestoacasa}, \hashtag{sciacalli} and \hashtag{italylockdown} as \textit{focal hashtags} to distinguish them from other hashtags.
We collected the tweets through \textit{Complex Science Consulting} ($@ComplexConsult$), which was authorised by Twitter, and used the \textit{ServiceConnect} crawler implemented in Mathematica 11.3. The collection of tweets comprises 39,943 tweets for \hashtag{iorestoacasa}, 26,999 for \hashtag{sciacalli} and 34,825 for \hashtag{italylockdown}. Retweets of the same text message were not considered. For each tweet, the language was detected. Pictures, links, and non-Italian content was discarded and stop-words (i.e. words without intrinsic meaning such as ``di''  (English: "of") and ``ma'' (English: "but") removed. Other interesting datasets with tweets about COVID-19 are available in \citep{chen2020covid,kleinberg2020measuring}.

\subsection{Multi-layer co-occurrence networks}

Word co-occurrence networks have been successfully used to characterise a wide variety of phenomena related to language acquisition and processing \cite{amancio2015probing,marinho2018labelled,vankrunkelsven2018predicting}. 
Recently, researchers have also used hashtags to investigate various aspects of social discourse. For instance, Stella et al. \cite{stella2018bots} showed that hashtag co-occurrence networks were able to characterise important differences in the social discourses promoted by opposing social groups during the Catalan referendum.
In this work we introduce MERCURIAL (Multi-layer Co-occurrence Networks for Emotional Profiling), a framework combining:
\begin{itemize}
    \item Hashtag co-occurrence networks (or hashtag networks) \cite{stella2018bots}. Nodes represent hashtags and links indicate the co-occurrence of any two nodes in the same tweet.
    \item Word co-occurrence networks (or word networks) \cite{amancio2015probing}. Nodes represent words and links represent the co-occurrence of any two words one after the other in a tweet without stop-words (i.e. words without an intrinsic meaning).
\end{itemize}{}
We combine these two types of networks in a multi-layer network to exploit the interdependence between hashtags and words. This new, resulting network enables us to contextualise hashtags, and capture their real meaning through context, thereby enhancing the accuracy of the emerging emotional profile. To build the multi-layer network, we first build the single hashtag and word layers. \begin{figure}[ht]
\centering
\includegraphics[width=15cm]{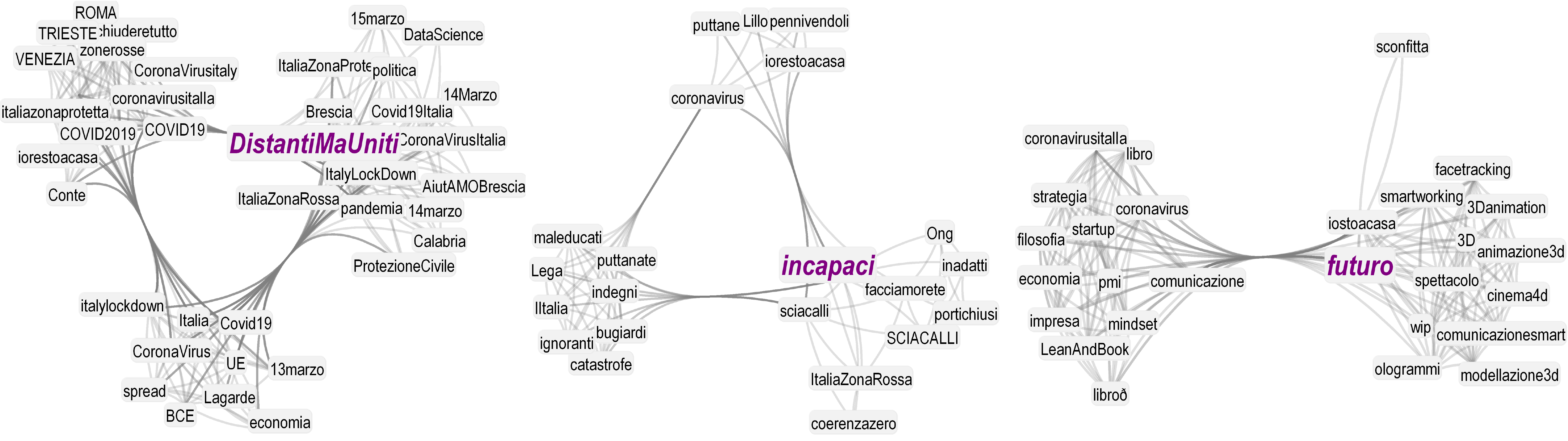}
\includegraphics[width=14cm]{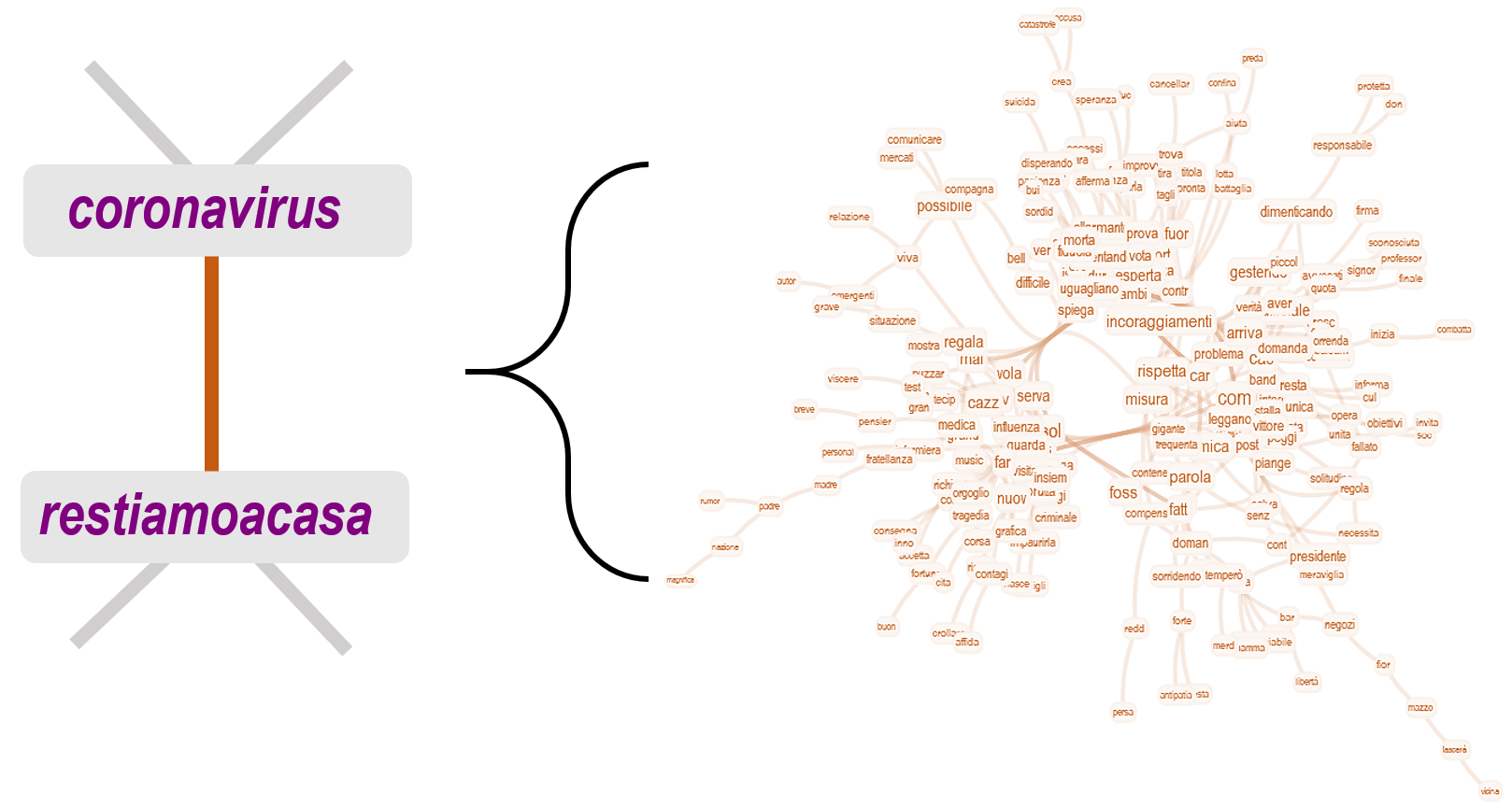}
\caption{\textbf{Top:} Example of co-occurrence networks for different hashtags: \hashtag{distantimauniti} (English: distant but united) in \hashtag{iorestoacasa} on the left, \hashtag{incapaci} (English: inept) in \hashtag{sciacalli} in the middle, and \hashtag{futuro} (English: future) in \hashtag{italylockdown} on the right. 
Clusters of co-occurring hashtags were obtained through spectral clustering \cite{newman2018networks}. These clusters highlight the co-occurrence of European-focused content, featuring hashtags like \hashtag{BCE} (i.e. European Central Bank), \hashtag{Lagarde} and \hashtag{spread} (i.e. spread between Italian and German bonds) together with social distance practices related to \hashtag{iorestoacasa}.  \textbf{Bottom:} In MERCURIAL, any link in a co-occurrence network of hashtags (left) corresponds to a collection of tweets whose words co-occur according to a word network (right). Larger words have a higher closeness centrality. }
\label{fig:1}
\end{figure}   For sake of simplicity, word networks are unweighted and undirected\footnote{Robustness checks were performed for various cut-offs in co-occurrence frequency, e.g. pruning co-occurrence networks according to co-occurrence weights did not change the most central hashtags.}. Note that the hashtag network was kept at a distinct level from word networks, e.g. common words were not explicitly linked with hashtags. As reported in Figure~\ref{fig:1}, each co-occurrence link between any two hashtags A and B (\hashtag{coronavirus} and \hashtag{restiamoacasa} in the figure) is relative to a word network, including all words co-occurring in all tweets featuring hashtags A and B.

The hashtag and word networks capture the co-occurrence of lexical entities within the structured online social discourse. Words possess meaning in language \cite{de2016large} and their network assembly is evidently a linguistic network. Similar to words in natural language, hashtags possess linguistic features that express a specific meaning and convey rich affect patterns \cite{stella2018bots}. The resulting networks capture the meaning of a collection of tweets by identifying which words/hashtags co-occurred together. 

This knowledge embedded in hashtag networks was used in order to identify the most relevant or central terms associated within a given collection of thematic tweets. Rather than using frequency to indicate centrality, which makes it difficult to compare hashtags that do not co-occur in the same message, the current work relies on distance-based measures to detect how central a hashtag is in the network.

The first measure that implements this notion is closeness centrality. Closeness $c(i)$ identifies how many links connect $i$ to all its neighbours and is formalised as follows:

\begin{equation}
c(i)=\frac{N}{\sum_{j=1}^{N} d_{ij}},
\end{equation}
where $d_{ij}$ is the network distance between $i$ and $j$, i.e. the smallest amount of links connecting nodes $i$ and $j$. In co-occurrence networks, nodes (i.e. hashtags and words) with a higher closeness tend to co-occur more often with each other or with other relevant nodes at short network distance. We expect that rankings of closeness centrality will reveal the most central hashtags in the networks for \hashtag{iorestoacasa}, \hashtag{sciacalli} and \hashtag{italylockdown}, in line with previous work in which closeness centrality was used to measure language acquisition and processing \cite{stella2019modelling,siew2019cognitive,stella2019innovation}. 

Importantly, closeness is a more comprehensive approach compared to the simpler frequency analysis. Imagine a collection of hashtags ${A,B,C,D,...}$. Computing the frequency of hashtag A co-occurring with hashtag B is informative about the frequency of the so-called 2-grams "AB" or "BA" but it does not consider how those hashtags co-occur with C, D, etc. In other words, a 2-gram captures the co-occurrence of two specific hashtags within tweets but does not provide the simultaneous structure of co-occurrences of all hashtags across tweets, for which a network of pairwise co-occurrences is required. On such a network, closeness can then highlight hashtags at short distance from all others, i.e. co-occurring in a number of contexts in the featured discourse. 

In addition to closeness, we also use \textit{graph distance entropy} to measure centrality. This centrality measure captures which hashtags are uniformly closer to all other hashtags in a connected network. Combining closeness with graph distance entropy led to successfully identifying words of relevance in conceptual networks with a few hundreds of nodes \cite{stella2019innovation}.

The main idea behind graph distance entropy is that it provides info about the spread of the distribution of network distances between nodes (i.e. shortest path), a statistical quantity that cannot be extracted from closeness (which is, conversely, a mean inverse distance). Considering the set $\textbf{d}^{(i)}\equiv(d_{i1},...,d_{ij},...,d_{iN})$ of distances between $i$ and any other node $j$ connected to it ($1 \leq j \leq N$) and $M_i=Max(\textbf{d}^{(i)})$, then graph distance entropy is defined as:

\begin{equation}
h(i)=-\frac{1}{\text{log}(M_i-1)} \sum_{k=1}^{M_i-1} p_{k}^{(i)}\text{log}p_{k}^{(i)},
\end{equation}

where $p_k$ is the probability of finding a distance equal to $k$. Therefore, $h(i)$ is a Shannon entropy of distances and it ranges between 0 and 1. In general, the lower the entropy, the more a node resembles a star centre \citep{stella2018distance} and is at equal distances from all other nodes. Thus, nodes with a lower $h(i)$ and a higher closeness are more uniformly close to all other connected nodes in a network. Words with \textit{simultaneously} low graph distance entropy and high closeness were found to be prominent words for early word learning \cite{stella2018distance} and mindset characterisation \cite{stella2019innovation}.

\subsection{Attributing meaning and emotions to focal hashtags by using word networks}

In addition to hashtag networks, we also build word networks obtained from a collection of tweets containing any combination of the focal hashtags \hashtag{iorestocasa} or \hashtag{sciacalli} and \hashtag{coronavirus}. For all tweets containing a given set of hashtags, we performed the following:

\begin{enumerate}
    \item Subdivide the tweet in sentences and delete all stop-words from each sentence, preserving the original ordering of words;
    \item Stem all the remaining words, i.e. identify the root or stem composing a given word. In a language such as Italian, in which there is a number of ways of adding suffixes to words, word stemming is essential in order to recognise the same word even when it is inflected for different gender, number or as a verb tense. For instance, \textit{abbandoneremo} (we will abandon) and \textit{abbandono} (abandon, abandonment) both represent the same stem \textit{abband};
    \item Draw links between a stemmed word and its subsequent one. Store the resulting edge list of word co-occurrences.
    \item Sentences containing a negation (i.e. "not") underwent an additional step parsing their syntactic structure. This was done in order to identify the target of negation (e.g. in "this is not peace", the negation refers to "peace"). Syntactic dependencies were not used for network construction but intervened in emotional profiling, instead (see below).
\end{enumerate}{}

The resulting word network also captures syntactic dependencies between words \cite{amancio2015probing} related by online users to a specific hashtag or combination of hashtag. We used closeness centrality to detect the relevance of words for a given hashtag. Text pre-processing such as word stemming and syntactic dependencies was performed using Mathematica 11.3, which was also used to extract networks and compute network metrics.

The presence of hashtags in word networks provided a way of linking words, which express common language, with hashtags, which express content but also summarise the topic of a tweet. Consequently, by using this new approach, the meaning attributed by users to hashtags can be inferred not only from hashtag co-occurrence but also from word networks. An example of MERCURIAL, featuring hashtag-hashtag and word-word co-occurrences, is reported in Figure 1 (bottom). In this example, hashtags \hashtag{coronavirus} and \hashtag{restiamoacasa} co-occurred together (left) in tweets featuring many co-occurring words (right). The resulting word network shows relevant concepts such as "incoraggiamenti" (English: encouragement) and "problemi" (English: problems), highlighting a positive attitude towards facing problems related to the pandemic. More in general, the attribution and reconstruction of such meaning was explored by considering conceptual relevance and emotional profiling in one or several word networks related to a given region of a hashtag co-occurrence network. 

\subsection{Emotional profiling}
As a first data source for emotional profiling, this work also used valence and arousal data from Warriner and colleagues \cite{warriner2013norms}, whose combination can reconstruct emotional states according to the well-studied circumplex model of affect \cite{russell1980circumplex,posner2005circumplex}. In psycholinguistics, word valence expresses how positively/negatively a concept is perceived (equivalently to sentiment in computer science). The second dimension, arousal, indicates the alertness or lethargy inspired by a concept. Having a high arousal and valence indicates excitement and joy, whereas a negative valence combined with a high arousal can result in anxiety and alarm \cite{posner2005circumplex}. Finally, some studies also include dominance or potency as a measure of the degree of control experienced \cite{warriner2013norms}. However, for reasons of conciseness, we focus on the two primary dimensions of affect: valence and arousal.

Going beyond the standard positive/negative/neutral sentiment intensity is of utmost importance for characterising the overall online perception of massive events \cite{ferrara2015quantifying}. Beyond the primary affective dimension of sentiment, the affect associated with current events \cite{stella2018bots} can also be described in terms of arousal \cite{unwin2002effects} and of basic emotions such as fear, disgust, anger, trust, joy, surprise, sadness, and anticipation. These emotions represent basic building blocks of many complex emotional states \cite{ekman1994nature}, and they are all self-explanatory except for anticipation, which indicates a projection into future events \cite{mohammad2010emotions}. Whereas fear, disgust, and anger (trust and joy) elicit negative (positive) feedback, surprise, sadness and anticipation have been recently evaluated as neutral emotions, including both positive and negative feedback reactions to events in the external world \cite{fonagy2018affect}.

To attribute emotions to individual words, we use the NRC lexicon \cite{mohammad2010emotions} and the circumplex model \cite{posner2005circumplex}. These two approaches allow us to quantify the \textit{emotional profile} of a set of words related to hashtags or combinations of hashtags. The NRC lexicon enlists words eliciting a given emotion. The circumplex model attributes valence and arousal scores to words, which in turn determine their closest emotional states.  Because datasets of similar size were not available for Italian, the data from the NRC lexicon and the Warriner norms were translated from English to Italian using a forward consensus translation of Google Translate, Microsoft Bing and DeepL translator, which was successfully used in previous investigations with Italian \cite{stella2020formab}. Although the valence of some concepts might change across languages \cite{warriner2013norms}, word stemming related several scores to the same stem, e.g. scores for "studio" (English: "study") and "studiare" (English: "to study") were averaged together and the average attributed to the stem root "stud". In this way, even if non-systematic cross-language valence shifting introduced inaccuracy in the score for one word (e.g. "studiare"), averaging over other words relative to the same stem reduced the influence of such inaccuracy. No statistically significant difference ($\alpha = 0.05$) was found between the emotional profiles of 200 Italian tweets, including 896 different stems, and their automatic translations in English, holding for each dimension separately (z-scores < 1.96).

Then, we build emotional profiles by considering the distribution of words eliciting a given emotion/valence/arousal and associated to specific hashtags in tweets. Assertive tweets with no negation were evaluated directly through a bag of words model, i.e. by directly considering the words composing them. Tweets including negations underwent an additional intermediate step where words syntactically linked to the negation were substituted with their antonyms \cite{miller1998wordnet} and then evaluated. Source-target syntactic dependencies were computed in Mathematica 11.3 and all words targeted by a negation word (i.e. \textit{no}, \textit{non} and \textit{nessuno} in Italian) underwent the substitution with their antonyms.

To determine whether the observed emotional intensity $r(i)$ of a given emotion in a set $S$ of words was compatible with random expectation, we perform a statistical test ($Z$-test) using the NRC dataset. Remember that emotional intensity here was measured in terms of richness or count of words eliciting a given emotion in a given network. As a null model, we use random samples as follows: let us denote by $m$ the number of words stemmed from $S$ that are also in the NRC dataset. Then, $m$ words from the NRC lexicon are sampled uniformly at random and their emotional profile is compared against that of the empirical sample. We repeated this random sampling 1000 times for each single empirical observed emotional profile $\{r(i)\}_i$. To ensure the resulting profiles are indeed compatible with a Gaussian distribution, we performed a Kolmogorov-Smirnov test ($\alpha=0.05$). All the tests we performed gave random distributions of emotional intensities compatible with a Gaussian distribution, characterised by a mean random intensity for emotion $i$, $r^{*}(i)$ and a standard deviation $\sigma^{*}(i)$. For each emotion, a z-score was computed:

\begin{equation}
    z=\frac{r(i) - r^{*}(i)}{\sigma^{*}(i)}.
\end{equation}{}

In the remainder of the manuscript, every emotional profile incompatible with random expectation was highlighted in black or marked with a check. Since we used a two-tailed $Z$-test (with a significance level of 0.05), this means that an emotional richness can either be higher or lower than random expectation.

\section{Results}

The investigated corpus of tweets represents a complex multilevel system, where conceptual knowledge and emotional perceptions are entwined on a number of levels. Tweets are made of text and include words, which convey meaning \cite{de2016large}. From the analysis of word networks, we can obtain information on the organisation of knowledge proper of social media users, which is embedded in their generated content \cite{amancio2015probing}. However, tweets also convey meaning through the use of hashtags, which can either refer to specific words or point to the overall topic of the whole tweet. Both words and hashtags can evoke emotions in different contexts, thus giving rise to complex patterns \cite{quercia2011our}. Similar to  words in natural language, the same hashtags can be perceived and used in language differently by different users, according to the context. 

The simultaneous presence of word- and hashtag-occurrences in tweets is representative of the knowledge shared by social media users when conveying specific content and ideas. This interconnected representation of knowledge can be exploited by simultaneously considering both hashtag-level and word-level information, since words specify the meaning attributed to hashtags. 

In this section we use MERCURIAL to analyse the data collected. We do so by characterising the hashtag networks, both in terms of meaning and emotional profiles. Precedence is given to hashtags as they not only convey meaning as individual linguistic units but also represent more general-level topics characterising the online discourse. Then, we inter-relate hashtag networks with word networks. Finally, we perform the emotional profiling of hashtags in specific contexts. The combination of word- and hashtag-networks specifies the perceptions embedded by online users around the same entities, e.g. coronavirus, in social discourses coming from different contexts. 


\subsection{Conceptual relevance in hashtag networks}

The largest connected components of the three hashtag networks included: 1000 hashtags and 8923 links for \hashtag{italylockdown}; 720 hashtags and 5915 links for \hashtag{sciacalli}; 6665 hashtags and 53395 links for \hashtag{italylockdown}. All three networks are found to be highly clustered (mean local clustering coefficient \cite{newman2018networks} of 0.82) and with an average distance between any two hashtags of 2.1. Only 126 hashtags were present in all the three networks. 

Table \ref{tab:1} reports the most central hashtags included in each corpus of tweets thematically revolving around \hashtag{iorestoacasa}, \hashtag{sciacalli} and \hashtag{italylockdown}. The ranking relies on closeness centrality, which in here quantifies the tendency for hashtags to co-occur with other hashtags expressing analogous concepts and, therefore, are at short network distance from each other (see Methods). Hence, hashtags with a higher closeness centrality represent the prominent concepts in the social discourse. This result is similar to those showing that closeness centrality captures concepts which are relevant for early word acquisition \cite{stella2019modelling} and production \cite{castro2019multiplex} in language. Additional evidence that closeness can capture semantically central concepts is represented by the closeness ranking, which assigns top-ranked positions to \hashtag{coronavirus} and \hashtag{COVID-19} in all three Twitter corpora. This is a consequence of the corpora being about the COVID-19 outbreak (and of the network metric being able to capture semantic relevance). 

In the hashtag network built around \hashtag{italylockdown}, the most central hashtags are relative to the coronavirus, including a mix of negative hashtags such as \hashtag{pandemia} (English: "pandemic") and positive ones such as \hashtag{italystaystrong}. Similarly, the hashtag network built around \hashtag{sciacalli} highlighted both positive (\hashtag{facciamorete} (English: "let's network") and negative (\hashtag{irresponsabili} -  English: "irresponsible") hashtags. However, the social discourse around \hashtag{sciacalli} also featured prominent hashtags from politics, including references to specific Italian politicians, to the Italian Government, and hashtags expressing protest and shame towards the acts of a prominent Italian politician. Conversely, the social discourse around \hashtag{iorestoacasa} included many positive hashtags, eliciting hope for a better future and the need to act responsibly (e.g. \hashtag{andratuttobene} - English: "everything will be fine", or \hashtag{restiamoacasa} - English: "let's stay at home"). The most prominent hashtags in each network (cf. Table 1) indicate the prevalence of a positive social discourse around \hashtag{iostoacasa} and the percolation of strong political debate in relation to the negative topics conveyed by \hashtag{sciacalli}. However, we want to extend these punctual observations of negative/positive valences of single hashtags to the overall global networks. To achieve this, we use emotional profiling. 

\subsection{Emotional profiling of hashtag networks}

Hashtags can be composed of individual or multiple words. By extracting individual words from the hashtags of a given network, it is possible to reconstruct the emotional profile of the social discourse around the focal hashtags \hashtag{sciacalli}, \hashtag{italylockdown} and \hashtag{iorestoacasa}. We tackle this by using the emotion-based \cite{mohammad2010emotions} and the dimension based \cite{posner2005circumplex} emotional profiles (see Methods).

\begin{table}[ht]
\caption{Top-ranked hashtags in co-occurrence networks based on closeness centrality. Higher ranked hashtags co-occurred with more topic-related concepts in the same tweet. In all three rankings, the most central hashtag was the one defining the topic (e.g. \hashtag{italylockdown}) and was omitted from the ranking.}
\centering
\centering
\label{tab:1}
\includegraphics[width=9.5cm]{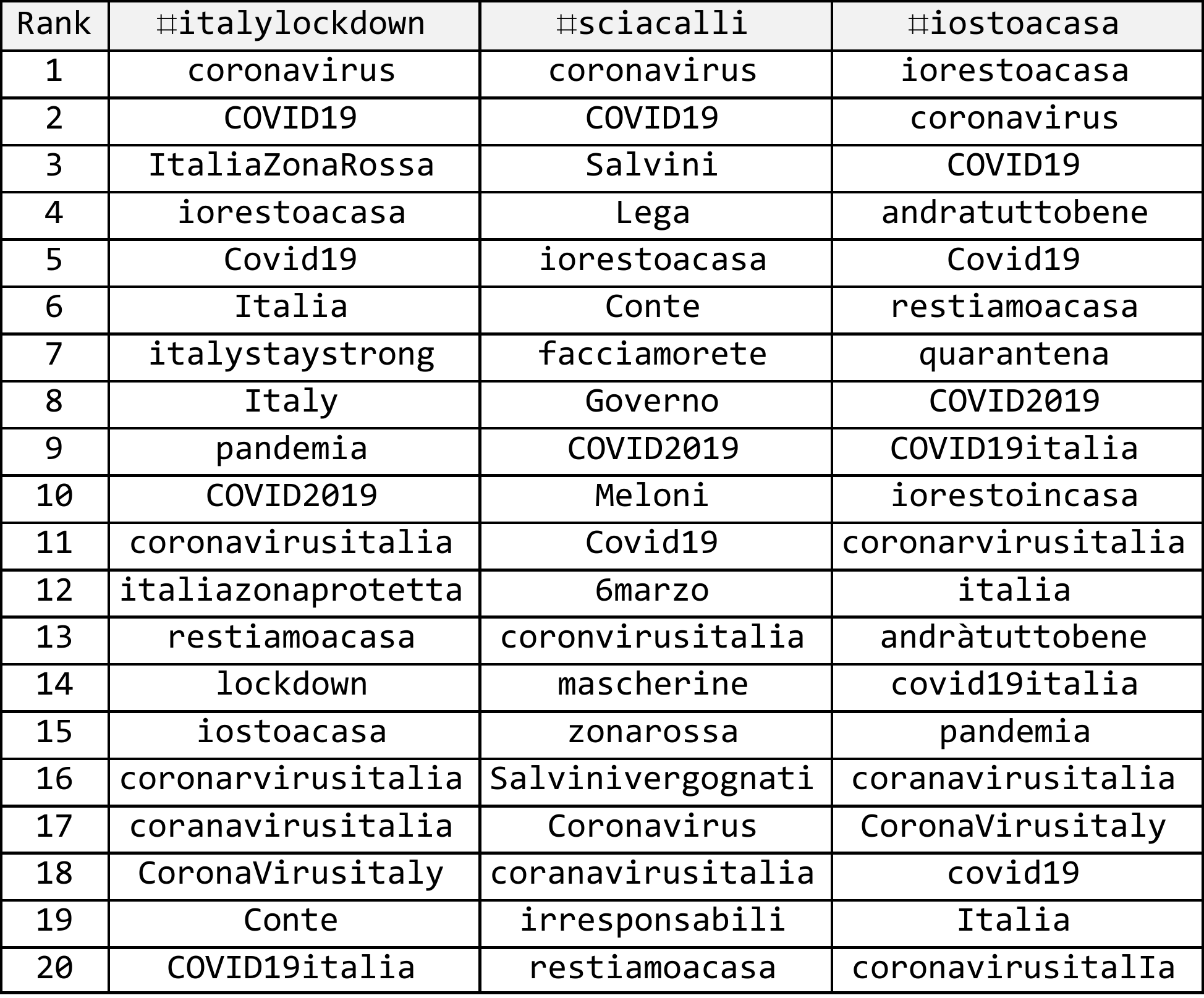}
\end{table}

The emotional profiles of hashtags featured in co-occurrence networks are reported in Figure \ref{fig:2} (top). The top section of the figure represents perceived valence and arousal represented as a circumplex model of affect \cite{posner2005circumplex}. This 2D space or disk is called \textit{emotional circumplex} and its coordinates represent emotional states that are well-supported by empirical behavioural data and brain research \cite{posner2005circumplex}. As explained also in the figure caption, each word is endowed with an $(x,y)$ coordinate expressing its perceived valence $(x)$ and arousal $(y)$. Different points indicate different emotional combinations. For instance, (1,0) is the point of maximum/positive valence and zero arousal, i.e. calmness; (0,-1) is the point of zero valence and minimum arousal, i.e. lethargy; (-0.6,+0.6) represents a point of strong negative valence and positive arousal, i.e. alarm. 

Figure \ref{fig:2}  reports the emotional profiles of all hashtags featured in co-occurrence networks for \hashtag{italylockdown} (left), \hashtag{sciacalli} (middle) and \hashtag{iorestoacasa} (right). To represent the interquartile range of all words for which valence/arousal rating are available, we use a neutrality range. Histograms falling outside of the neutrality range indicate specific emotional states expressed by words included within hashtags (e.g. \hashtag{pandemia} contains the word "pandemia" with negative valence and high arousal).

\subsubsection{Emotional profiling of hashtag networks through the circumplex model}
\begin{figure}[ht]
\centering
\includegraphics[width=16cm]{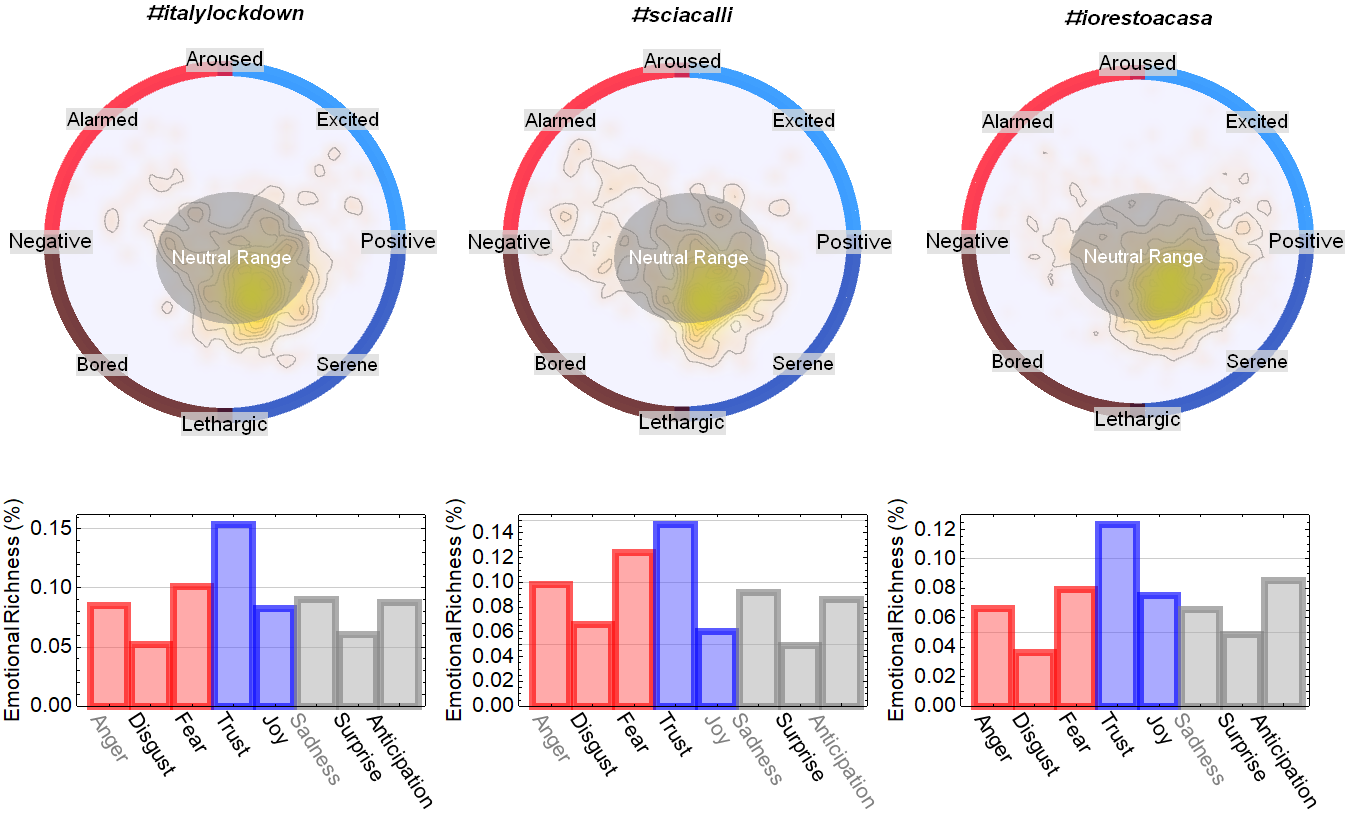}
\caption{Emotional profiles of all hashtags featured in co-occurrence networks for \hashtag{italylockdown}  (left), \hashtag{sciacalli} (middle) and \hashtag{iorestoacasa}  (right). \textbf{Top:} Circumplex emotional profiling. All hashtags representing one or more words were considered. For each word, valence ($x$-coordinate) and arousal ($y$- coordinate) scores were attributed (see Methods) resulting in a 2D density histogram (yellow overlay) relative to the probability of finding an hashtag in a given location in the circumplex, the higher the probability the stronger the colour. Regions with the same probabilities are enclosed in grey lines. A neutrality range indicates where 50\% of the words in the underlying valence/arousal dataset would fall and it thus serves as a reference value for detecting abnormal emotional profiles. Distributions falling outside of this range indicate deviations from the median behaviour (i.e. interquartile range, see Methods). \textbf{Bottom:} NRC-based emotional profiling, detecting how many hashtags inspired a given emotion in a hashtag network. Results are normalised over the total number of hashtags in a networks. Emotions compatible with random expectation were highlighted in gray.}
\label{fig:2}
\end{figure}

In Figure \ref{fig:2} (left, top), the peak of the emotional distribution for hashtags associated with \hashtag{italylockdown} falls within the neutrality range. This finding indicates that hashtags co-occurring with \hashtag{italylockdown}, a neutral hashtag by itself, were also mostly emotionally neutral conceptual entities. Despite this main trend, the distribution also features deviations from the peak mostly in the areas of calmness and tranquillity (positive valence, lower arousal) and excitement (positive valence, higher arousal). Weaker deviations (closer to the neutrality range) were present also in the area of anxiety.

This reconstructed emotional profile indicates that the Italian social discourse featuring  \hashtag{italylockdown} was mostly calm and quiet, perceiving the lockdown as a positive measure for countering responsibly the COVID-19 outbreak.
 
Not surprisingly, the social discourse around \hashtag{sciacalli} shows a less prominent positive emotional profile, with a higher probability of featuring hashtags eliciting anxiety, negative valence and increased states of arousal, as it can be seen in Figure \ref{fig:2} (center, top). This polarised emotional profile represents quantitative evidence for the coexistence of mildly positive and strongly negative content within the online discourse labelled by \hashtag{sciacalli}. This is further evidence that the negative hashtag \hashtag{sciacalli} was indeed used by Italian users to denounce or raise alarm over the negative implications of the lockdown, especially in relation to politics and politicians' actions. However, the polarisation of political content and debate over social media platforms has been encountered in many other studies \cite{brito2020complex,bail2018exposure,stella2018bots} and cannot be attributed to the COVID-19 outbreak only. 

Finally, Figure \ref{fig:2} (top right) shows that positive perception was more prominently reflected in the emotional profile of \hashtag{iorestoacasa}, which was the hashtag massively promoted by the Italian Government for supporting the introduction of the nationwide lockdown in Italy. The emotional profile of the 6000 hashtags co-occurring with \hashtag{iorestoacasa} indicate a considerably positive and calm perception of domestic confinement, seen as a positive tool to stay safe and healthy. The prominence of hopeful hashtags in association with \hashtag{iorestoacasa}, as reported in the previous subsection, indicate that many Italian Twitter users were serene and hopeful about staying at home at the start of lockdown.

\subsubsection{Emotional profiling of hashtag networks through basic emotions}

Hashtag networks were emotionally profiled not only by using the circumplex model (see above) but also by using basic emotional associations taken from the NRC Emotion lexicon (Figure \ref{fig:2}, bottom). Across all hashtag networks, we find a statistically significant peak \footnote{The computed z-scores were: 2.89 > 1.96 (\hashtag{italylockdown}), 2.79 > 1.96 (\hashtag{sciacalli}) and 2.66 > 1.96 (\hashtag{iorestoacasa}).} in trust, analogous of the peaks close to emotions of calmness and serenity an observed in the circumplex models. However, all the hashtag networks included also negative emotions like anger and fear, which are natural human responses to unknown threats and were observed also with the circumplex representations. The intensity of fearful, alarming and angry emotions is stronger in the \hashtag{sciacalli} hashtag network, which was used by social users to denounce, complain and express alertness about the consequences of the lockdown. 

In addition to the politically-focused jargon highlighted by closeness centrality alone, by combining closeness with graph distance entropy (see Methods and \cite{stella2019innovation}) we identify other topics which are uniformly at short distance from others in the social discourse around \hashtag{sciacalli}, such as: \hashtag{mascherine} (English: "protective masks", which was also ranked high by using closeness only), \hashtag{amuchina} (the most popular brand, and synonym of, hand sanitiser), \hashtag{supermercati} (English: "supermarkets"). This result suggests an interesting interpretation of the negative emotions around \hashtag{sciacalli}. Beside the inflaming political debate and the fear of the health emergency, in fact, a \textit{third} element emerges: Italian twitter users feared and were angry about the raiding and stockpiling of first aid items, symptoms of panic-buying in the wake of the lockdown.
 
\subsection{Assessing conceptual relevance and emotional profiles of hashtags via word networks} 

The above comparisons indicate consistency between dimension-based (i.e. the circumplex) and emotion-specific emotional profiling. Since the latter offers also a more precise categorisation of words in emotions, we will focus on emotion-specific profiling.
Importantly, to fully understand the emotional profiles outlined above, it is necessary to identify the language expressed in tweets using a given combination of hashtags (see also Figure \ref{fig:1}, bottom). As the next step of the MERCURIAL analysis, we gather all tweets featuring the focal hashtags \hashtag{italylockdown},  \hashtag{sciacalli}, or \hashtag{iorestoacasa} and any of their co-occurring hashtags and build the corresponding word networks, as explained in the Methods. Closeness centrality over these networks provided the relevance of each single word in the social discourse around the topic identified by a hashtag. Only words with closeness higher than the median were reported.

\begin{figure}[ht]
\centering
\includegraphics[width=15cm]{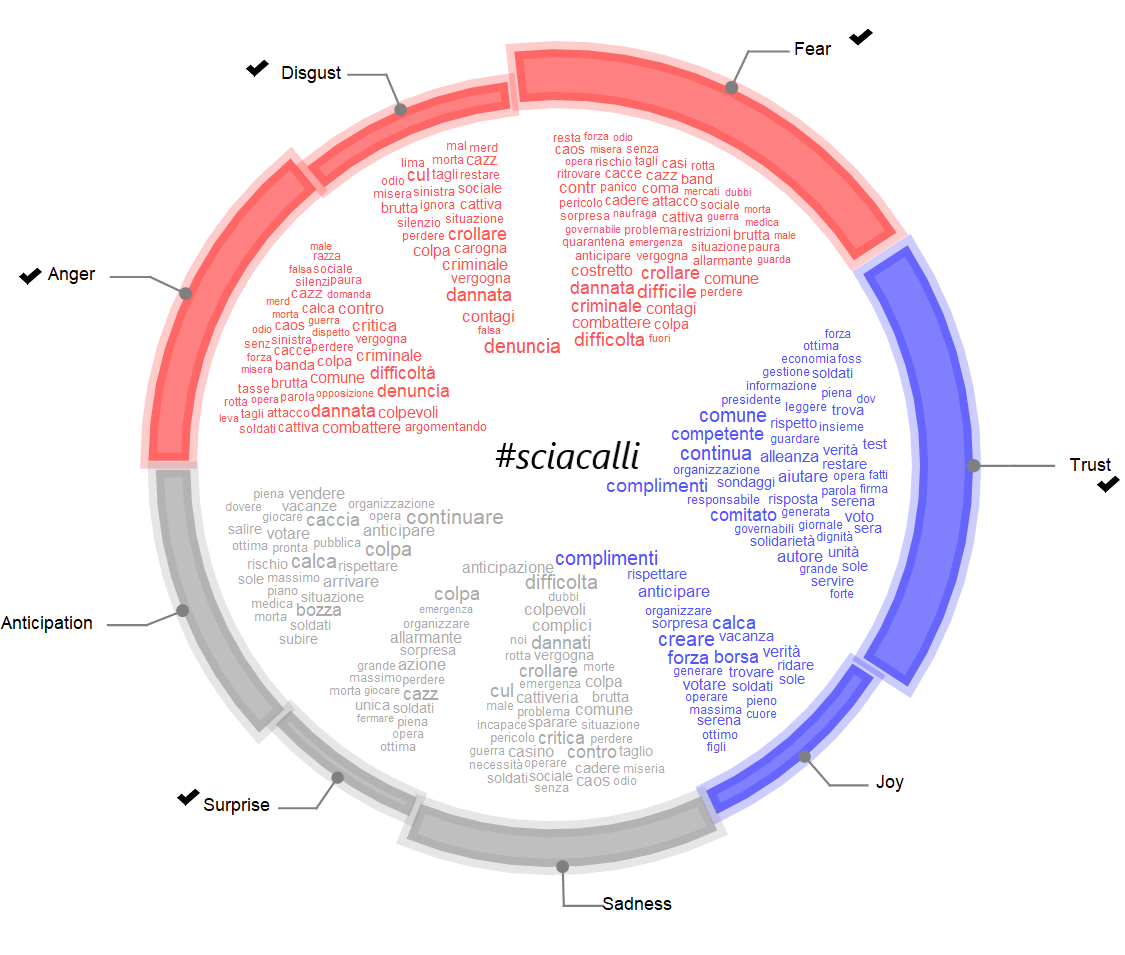}
\caption{Emotional profile and word cloud of the language used in tweets with \hashtag{sciacalli}. Words are organised according to the emotion they evoke. Font size is larger for words of higher closeness centrality in the word co-occurrence network relative to the hashtag (see Methods). Every emotion level incompatible with random expectation is highlighted with a check mark.}
\label{fig:3}
\end{figure}

Figure \ref{fig:3} shows the cloud of words appearing in all tweets that include \hashtag{sciacalli}, displayed according to their NRC emotional profile. Similar to the emotional profile extracted from hashtags co-occurring with \hashtag{sciacalli}, the words used in tweets with this hashtag also display a polarised emotional profile with high levels of fear and trust. Thanks to the multi-layer analysis, this dichotomy can now be better understood in terms of the individual concepts eliciting it.

By using closeness on word networks, we identified concepts such as "competente" (English: "competent"), "continua" (English: "continue", "keep going"), and "comitato" (English: "committee") to be relevant for the trust-sphere. These words convey trust in the expert committees appointed by the Italian Government to face the pandemic and protect the citizens. We find that other prominent words contributing to make the discourse around \hashtag{sciacalli} trustful are "aiutare" (English: "to help"), "serena" (English: "serene"), "rispetto" (English: "respect") and "verità" (English: "truth"), which further validate a trustful, open-minded and fair perception of the political and emergency debate outlined above. This perception was mixed with negative elements, mainly eliciting fear but also sadness and anger. The jargon of a political debate emerges in the word cloud of fear: "difficoltà" (English: "difficulty"), "criminale" (English: "criminal"), "dannati" (English: "scoundrel"), "crollare" (English: "to break down"), "banda" (English: "gang"), "panico" (English: "panic") and "caos" (English: "chaos"). These words indicate that Twitter users felt fear directed to specific targets. A speculative explanation for exorcising fear can be finding a scapegoat and then target it with anger. The word cloud of such emotion supports the occurrence of such phenomenon by featuring words like "denuncia" (English: "denouncement"), "colpevoli" (English: "guilty"), "vergogna" (English: "shame"), "combattere" (English: "to fight") and "colpa" (English: "blame"). The above words are reflected also in other emotions like sadness, which features also words like "cadere" (English: "to fall") and "miseria" (English: "misery", "out of grace"). 

These prominent words in the polarised emotional profile of \hashtag{sciacalli}, suggest that Twitter users feared criminal behaviour, possibly related to unwise political debates or improper stockpiling of supplies (as showed by the hashtag analysis). Our findings also suggest that the reaction to such fearful state, which also projects sadness about negative economic repercussions, was split into a strong, angry denounce of criminal behaviour and messages of trust for the order promoted by competent organisations and committees. It is interesting to note that, according to Ekman's theory of basic emotions \cite{ekman1994nature}, a combination of sadness and fear can be symptomatic of desperation, which is a critical emotional state for people in the midst of a pandemic-induced lockdown.

\begin{figure}[ht!]
\centering
\includegraphics[width=12.2cm]{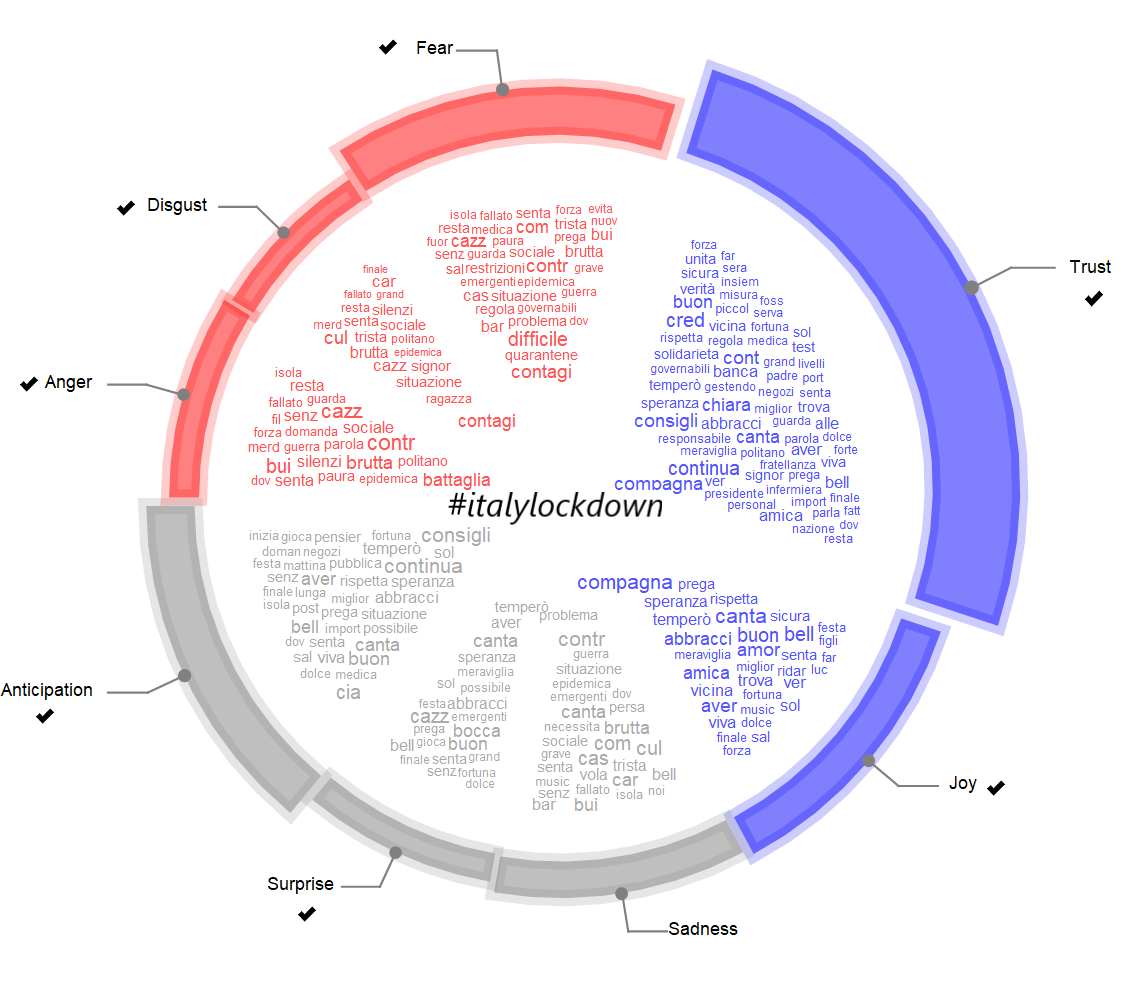}
\includegraphics[width=12.2cm]{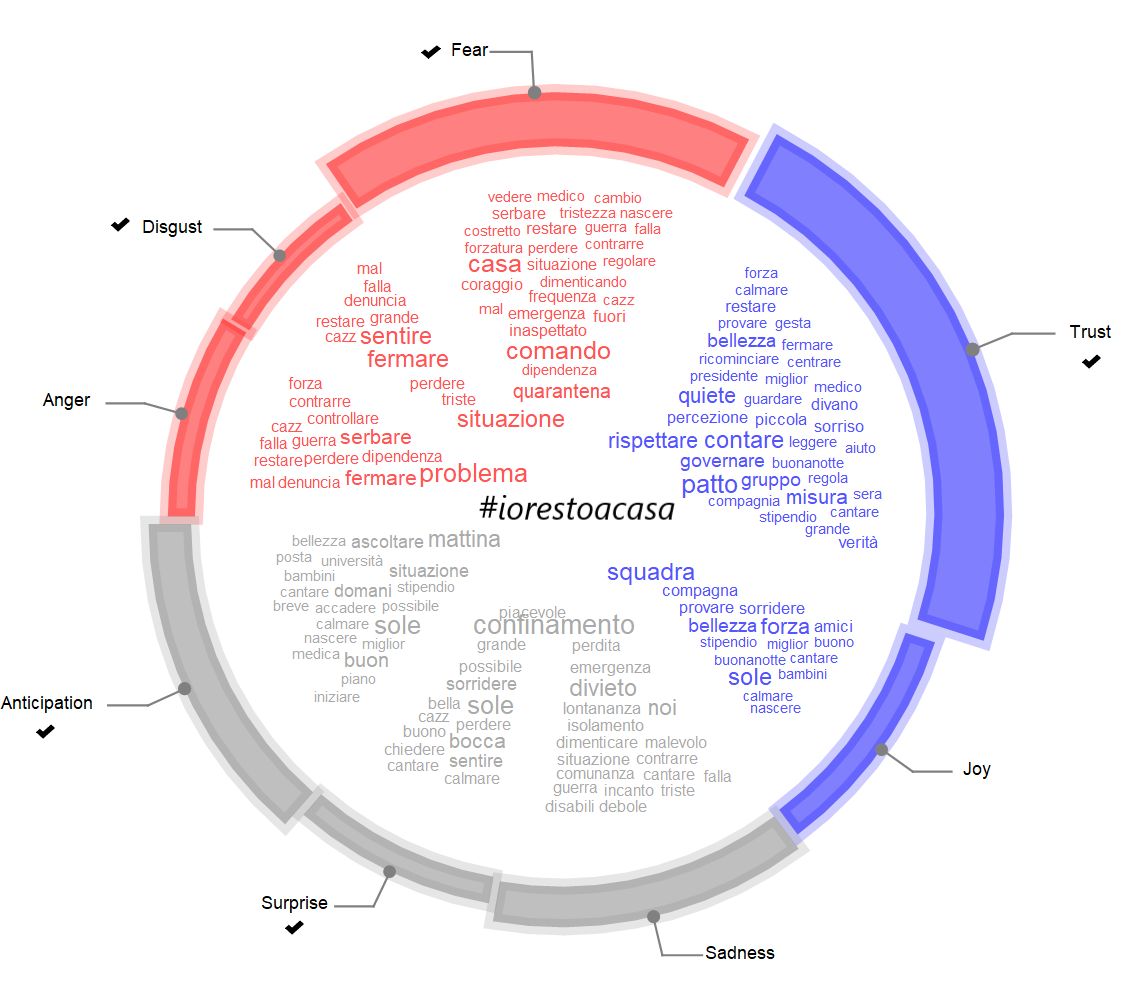}
\caption{Emotional profile and word cloud of the language used in tweets with \hashtag{italylockdown}  (top) and \hashtag{iorestoacasa}  (bottom). Words are organised according to the emotion they evoke. Font size is larger for words of higher closeness centrality in the word co-occurrence network relative to the hashtag (see Methods). Every emotional richness incompatible with random expectation is highlighted with a check mark.}
\label{fig:4}
\end{figure}

The same analysis is reported in Figure \ref{fig:4} for the social discourse of \hashtag{italylockdown} (top) and \hashtag{iorestoacasa} (bottom). In agreement with the circumplex profiling, for both \hashtag{italylockdown} and \hashtag{iorestoacasa} the intensity of fear is considerably lower than trust.

However, when investigated in conjunction with words, the overall emotional profile of \hashtag{italylockdown} appears to be more positive, displaying higher trust and joy and lower sadness, than the emotional profile of \hashtag{iorestoacasa}. Although the difference is small, this suggests that hashtags alone are not enough to fully characterise the perception of a conceptual unit, and should always be analysed together with the natural language associated to them.

The trust around \hashtag{italylockdown}  comes from concepts like "consigli" (English: "tips", "advice"), "compagna" (English: "companion", "partner"), "chiara" (English: "clear"), "abbracci" (English:"hugs") and "canta" (English: "sing"). These words and the positive emotions they elicit suggest that Italian users reacted to the early stages of the lockdown with a pervasive sense of commonality and companionship, reacting to the pandemic with externalisations of positive outlooks for the future, e.g. by playing music on the balconies\footnote{This phenomenon was also mimicked in other countries later, and extensively reported by traditional media, see https://tinyurl.com/balconicovid , Last Access: 20/04/2020 )}.

Interestingly, this positive perception co-existed with a more complex and nuanced one. Despite the overall positive reaction, in fact, the discourse on \hashtag{italylockdown} also shows fear for the difficult times facing the contagion ("contagi") and the lockdown restrictions ("restrizioni"), and also anger, identifying the current situation as a fierce battle ("battaglia") against the virus. 

The analysis of anticipation, the emotional state projecting desires and beliefs into the future, shows the emergence of concepts such as "speranza" (English: "hope"), "possibile" (English: "possible") and "domani" (English: "tomorrow"), suggesting a hopeful attitude towards a better future. 

The social discourse around \hashtag{iorestoacasa} brought to light a similar emotional profile, with a slightly higher fear towards being quarantined at home (quarantena (English: "quarantine"), comando (English: "command", "order", emergenza (English: "emergency"). Both surprise and sadness were elicited by the the word "confinamento" (English: "confinement"), which was prominently featured in the network structure arising from the tweets we analysed. 

In summary, the above emotional profiles of hashtags and words from the 101,767 tweets suggest that Italians reacted to the lockdown measure with: 
\begin{enumerate}
    \item a fearful denounce of criminal acts with political nuances and sadness/desperation about negative economic repercussions (from \hashtag{sciacalli});
    \item positive and trustful externalisations of fraternity and affect, combined with hopeful attitudes towards a better future (from \hashtag{italylockdown} and \hashtag{iorestoacasa});
    \item a mournful concern about the psychological weight of being confined at home, inspiring sadness and disgust towards the health emergency (from \hashtag{iorestoacasa}).
\end{enumerate}{}

\subsection{Hashtag co-occurrence contextually influences hashtag emotional profiles}
 
\begin{figure}[ht!]
\centering
\includegraphics[width=11.5cm]{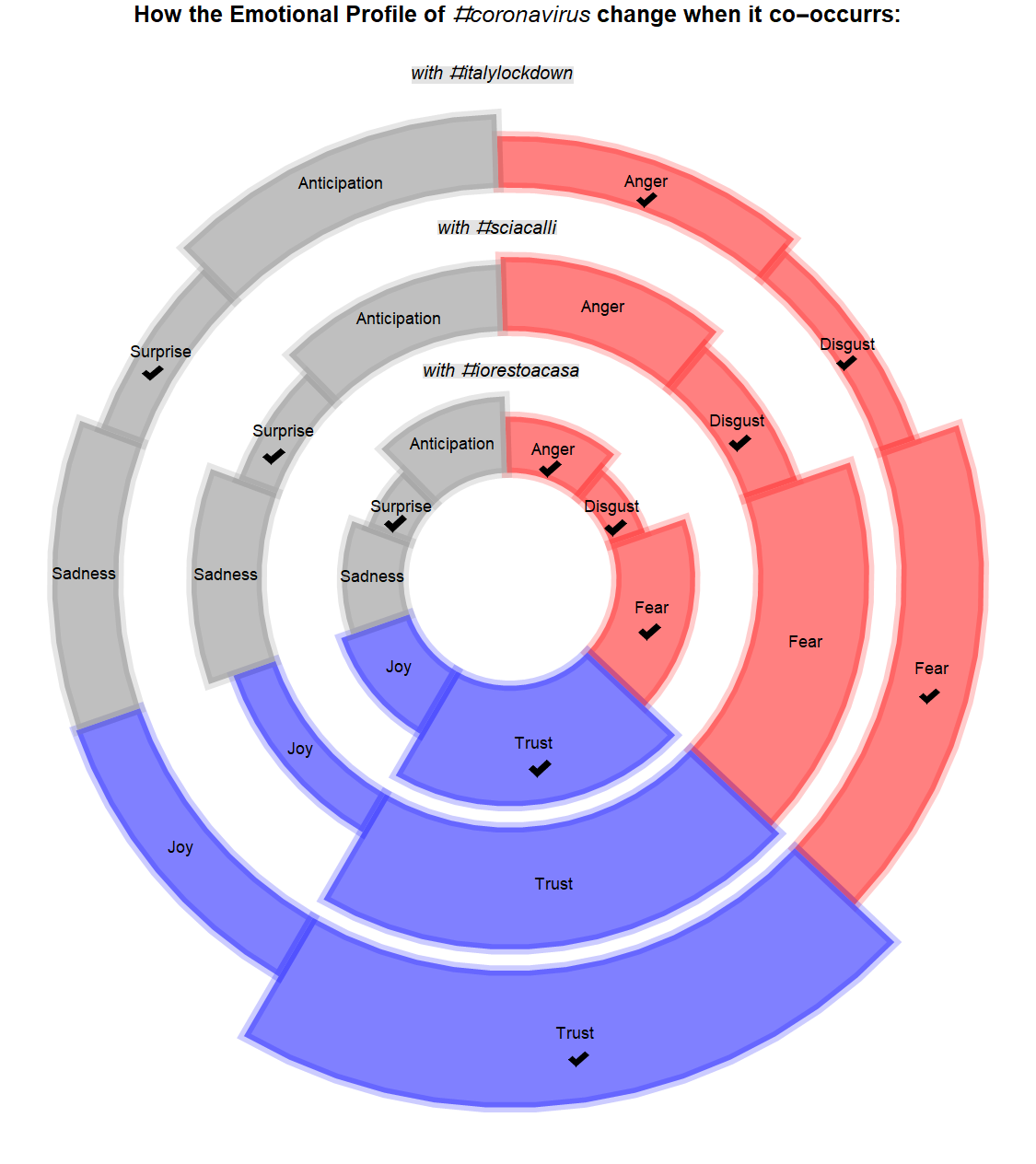}
\includegraphics[width=4.cm]{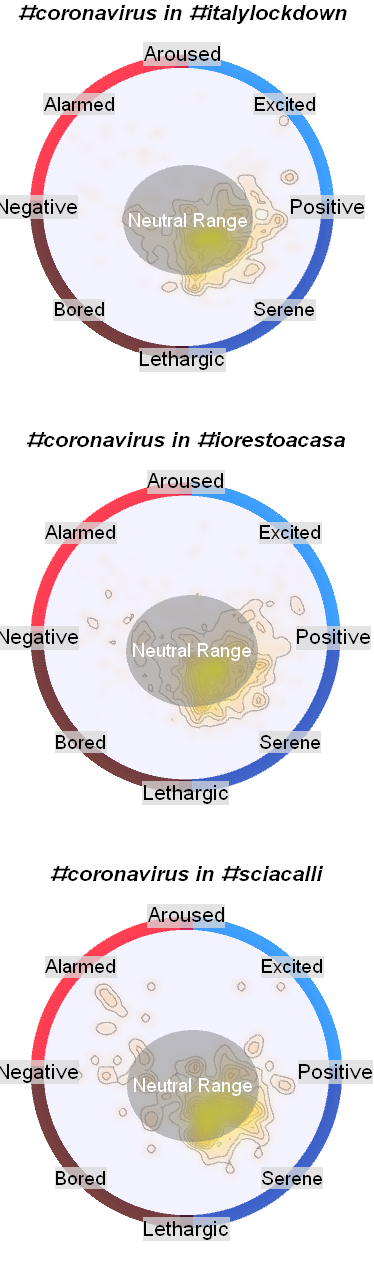}
\caption{\textbf{Left:} Emotional profiles with the NRC lexicon of the words occurring in tweets with the hashtag \hashtag{coronavirus}  when it co-occurs with \hashtag{italylockdown}  (outer circle), \hashtag{sciacalli} (middle circle) and \hashtag{iorestoacasa}  (inner circle). \textbf{Right:} The same emotional profiles detected through the valence/arousal emotional circumplex.  Every emotional richness incompatible with random expectation is highlighted with a check mark.}
\label{fig:5}
\end{figure}

In the previous section we showed our findings on how Italians perceived the early days of lockdown on social media. But what about their perception of the ultimate cause of such lockdown, COVID-19? To better reconstruct the perception of \hashtag{coronavirus}, it is necessary to consider the different contexts where this hashtag occurs. Figure \ref{fig:5} displays the reconstruction of the emotional profile of words used in tweets with \hashtag{coronavirus} and either \hashtag{italylockdown}, \hashtag{sciacalli}, or \hashtag{iorestoacasa}. 

Our results suggest that the emotional profiles of language used in these three categories of tweets are different. For example, when considering tweets including \hashtag{sciacalli}, which the previous analysis revealed being influenced by political and social denounces of criminal acts, \hashtag{coronavirus} is perceived with a more polarised fear/trust dichotomy.
Although \hashtag{coronavirus} was perceived as trustful as random expectations when co-occurring with \hashtag{sciacalli} (z-score: 1.69 < 1.96), it was perceived with significantly higher trust when appearing in tweets with \hashtag{iorestoacasa} (z-score: 3.05 > 1.96) and \hashtag{italylockdown} (z-score: 3.51 > 1.96). To reinforce this picture, the intensity of fear towards \hashtag{coronavirus} was statistically significantly lower than random expectation in the discourse of \hashtag{iorestoacasa} (z-score: -2.35 < -1.96) and \hashtag{italylockdown} (z-score: -3.01 < -1.96).

This difference is prominently reflected in both the circumplex model (Figure \ref{fig:5}, right) and the NRC emotional profile (Figure \ref{fig:5}, left), although in the latter both emotional intensities are compatible with random expectation. 
These quantitative comparisons provide data-driven evidence that Twitter users perceived the same conceptual entity, i.e. COVID-19, with a higher trust when associating it to concrete means for hampering pathogen diffusion like lockdown and house confinement, and with a higher fear when denouncing the politics and economics behind the pandemic.

However, social distancing, lockdown and house confinement clearly do not have only positive sides. Rather, as suggested by our analysis, they bear complex emotional profiles, where sadness, anger and fear towards the current situation and future developments have been prominently expressed by Italians on social media.

\section{Discussion}

This study delved into the massive information flow of Italian social media users in reaction to the declaration of the pandemic status of COVID-19 by WHO, and the announcement of the nationwide lockdown by the Italian Government in the first half of March 2020. We explored the emotional profiles of Italians during this period by analysing the social discourse around the official lockdown hashtag promoted by the Italian Government (\hashtag{iorestoacasa}), together with a most trending hashtag of social protest (\hashtag{sciacalli}), and a general hashtag about the lockdown (\hashtag{italylockdown}).

The fundamental premise of this work is that social media opens a window on the minds of millions of people \cite{quercia2011our}. Monitoring social discourse on online platforms provides unprecedented opportunities for understanding how different categories of people react to real world events \cite{ferrara2015quantifying,stella2018bots,stella2020text}.

\subsection{Impact of cognitive network science for social media analysis}

Here we introduced a new framework, Multi-layer Co-occurrence Networks for Emotional Profiling (MERCURIAL), which is based on cognitive network science and that allowed us to: 

(i) quantitatively structure social discourse as a multi-layer network of hashtag-hashtag and word-word co-occurrences in tweets; 

(ii) identify prominent discourse topics through network metrics backed up by cognitive interpretation \cite{siew2019cognitive}; 

(iii) reconstruct and cross-validate the emotional profile attributed to each hashtag or topic of conversation through the emotion lexicon and the circumplex model of affect from social psychology and cognitive neuroscience \cite{posner2005circumplex}. 

Our interdisciplinary framework provides a first step in combining network and cognitive science principles to quantify sentiment for specific topics. Our analysis also included extensive robustness checks (e.g. selecting words based on different centrality measures, statistical testing for emotions), further highlighting the potential of the framework.

The analysis of concept network centrality identified hashtags of political denounce and protest against irrational panic buying (e.g. face masks and hand sanitiser) around \hashtag{sciacalli} but not in the hashtag networks for \hashtag{italylockdown} and \hashtag{iorestoacasa}. Our results also suggest that the social discourse around \hashtag{sciacalli} was further characterised by fear, anger, and trust, whose emotional intensity was significantly stronger than random expectation. We also found that the most prominent concepts eliciting these emotions revolve around social denounce (anger), concern for the collective well-being (fear), and the measures implemented by expert committees and authorities (hope). This interpretation is supported also by Plutchik's wheel of emotions \cite{plutchik1991emotions}, according to which combinations of anger, disgust and anticipation can be symptoms of aggressiveness and contempt. However, within Plutchik's wheel, trust and fear are not in direct opposition.

\subsection{Evidence for emotional polarisation around COVID-19}

The polarisation of positive/negative emotions observed around \hashtag{sciacalli} might be a direct consequence of a polarisation of different social users with heterogeneous beliefs, which is a phenomenon present in many social systems \cite{brito2020complex} but is also strongly present in social media through the creation of echo chambers enforcing specific narratives and discouraging the discussion of opposing views \cite{bail2018exposure,cinelli2020covid,brugnoli2019recursive,ciulla2012beating,davis2020phase}. 

Emotional polarisation might therefore be a symptom of a severe lack of social consensus across Italian users in the early stages of the lockdown induced by COVID-19. In social psychology, social consensus is a self-built perception that the beliefs, feelings, and actions of others are analogous to one's own \cite{krueger1998perception}. Destabilising this perception can have detrimental effects such as reducing social commitment towards public good or even lead to a distorted perception of society, favouring self-distrust and even conditions such as social anxiety \cite{krueger1998perception}. Instead, acts such as singing from the balconies together can reduce fear and enhance self-trust \cite{unwin2002effects}, as well as promote commitment and social bonding \cite{pearce2015ice}, which is also an evolutionary response to help coping with a threat, in this case a pandemic, through social consensus. When interpreted under the lens of social psychology, the flash mobs documented by traditional media and identified here as relevant by semantic network analysis for \hashtag{italylockdown} and \hashtag{iorestoacasa} become important means of facing the distress induced by confinement \cite{krueger1998perception,unwin2002effects,pearce2015ice}.

\subsection{Implications of the detected anger and fear over self-awareness and violence}

Anger and fear permeated not only \hashtag{sciacalli} but were found, to a lesser extent, also in association with other hashtags such as \hashtag{iorestoacasa} or \hashtag{italylockdown}. Recent studies (cf. \cite{christensen2019just}) found that anger and fear can drastically reduce individuals' sense of agency, a subjective experience of being in control of our own actions, linking this behavioural/emotional pattern also to alteration in brain states. In turn, a reduced sense of agency can lead to losing control, potentially committing violent, irrational acts \cite{christensen2019just}. Consequently, the strong signals of anger and fear detected here represent red flags about a building tension manifested by social users which might contribute to the outbreak of violent acts or end up in serious psychological distress due to lowered self-control.

One of the most direct implications of the detected strong signals of fear, anger and sadness is represented by increased violent behaviour. In cognitive psychology, the General Aggression Model (GAM) \cite{dewall2011general} is a well-studied model for predicting and understanding violent behaviour as the outcome of a variety of factors, including personality, situational context and the personal internal state of emotion and knowledge. According to GAM, feeling emotions of anger in a situation of confinement can strongly promote violent behaviour. In Italy, the emotions of anger and anxiety we detected through social media are well reflected in the dramatic rise in reported cases of domestic violence. For instance, the anti-violence centers of D.i.Re (\textit{Donne in Rete Contro la Violenza}) reported an anomalous increase of +74.5\% in the number of women looking for help for domestic violence in March 2020 in Italy\footnote{(see the official report in Italian https://tinyurl.com/direcontrolaviolenza , Last Access: 20/04/2020)}. Hence, monitoring social media can be insightful about potential tensions mediated and discussed by large populations, a topic in need for further research and with practical prominent repercussions for fighting COVID-19.

\subsection{Contextual shifts in emotions around the novel coronavirus}

As discussed, we found the hashtag \hashtag{coronavirus} to be central across all considered hashtag networks. However, our analysis outlined different emotional nuances of  \hashtag{coronavirus} across different networks. In psycholinguistics, contextual valence shifting \cite{polanyi2006contextual} is a well-known phenomenon whereby the very same conceptual unit can be perceived wildly differently by people according to its context. This phenomenon suggests the importance of considering words in a contextual manner, by comparison to each other, as it was performed in this study, rather than alone. Indeed, contexts can change the meaning and emotional perception of many words in language. We showed here that the same connotation shifting phenomenon \cite{polanyi2006contextual} can happen also for hashtags. 
Online users perceived \hashtag{coronavirus}  with stronger intensities of trust and lower fear (than random expectation) when using that hashtag in the context of \hashtag{iorestoacasa} and \hashtag{italylockdown}, but not when associated to \hashtag{sciacalli}. This shifting underlines the importance of considering contextual information surrounding a hashtag in order to better interpret its nuanced perception. To this aim, cognitive networks represent a powerful tool, providing quantitative metrics (such as graph distance entropy) that would be otherwise not applicable with mainstream frequency approaches in psycholinguistics. 

\subsection{Limitations and future research}

MERCURIAL facilitates a quantitative characterisation of the emotions attributed to hashtags and discourses. Nonetheless, it is important to bear in mind that the analysis we conducted relies on some assumptions and limitations. For instance, following previous work \cite{stella2018bots}, we built unweighted and undirected networks, neglecting information on how many times hashtags co-occurred. Including these weights would be important for detecting communities of hashtags, beyond network centrality. Notice that including weights would come at the cost of not being able to use graph distance entropy, which is defined over unweighted networks and was successfully used here for exposing the denounce of panic buying in \hashtag{sciacalli}. Another limitation is relative to the emotional profiling performed with the NRC lexicon, in which the same word can elicit multiple emotions. Since we measured emotional intensity by counting words eliciting a given emotion (plus the negations, see Methods), a consequence was the repetition of the same words across the sectors of the above word clouds. Building or exploiting additional data about the predominance of a word in a given emotion would enable us to identify words which are peripheral to a given emotion, reduce repetitions and offer even more detailed emotional profiles. Recently, forma mentis networks \cite{stella2020forma,stella2019innovation} have been introduced as a method to detect the organisation of positive/negative words in the mindsets of different individuals. A similar approach might be followed for emotions in future research. Acting upon specific emotions rather than using the circumplex model would also solve another problem, in that the attribution of arousal to individual words is prone to more noise, even in mega-studies, compared to detecting word valence \cite{vad-acl2018}. Another limitation is that emotional profiles might fluctuate over time. The insightful results outlined and discussed here were aggregated over a short time window, thus reducing the impact of aggregation itself. Future analyses on longer time windows should adopt time-series for investigating emotional patterns, addressing key issues like non-stationary tweeting patterns over time and statistical scarcity due to tweet crawling (see also \cite{stella2018bots}).

The current analysis has focused on aggregated tweets, but previous studies have shown both stable individual and intercultural differences in affect \cite{kuppens2017relation}, especially for dimensions such as arousal. Similarly, some emotions are harder to measure than others, which might affect reliability and thus underestimate their contribution. The current approach estimates emotional profiles on the basis of a large set of words, which will reduce some language-specific differences. The collection of currently missing large-scale Italian normative datasets for lexical sentiment could further improve the accuracy of the findings. 
This study approaches the relation between emotions and mental distress mostly from the perspective that attitudes and emotions of the author are conveyed in the linguistic content. 
However, the emotion profile might also have implications for readers as well, as recent research suggests that even just reading words of strong valence/arousal can have deep somatic and visceral effects, e.g. raising heart beat or promoting involuntary muscle tension \cite{vergallito2019somatic}. 
Furthermore, authors and readers participate in an information network, and quantifying which tweets are liked or retweeted depending on the structure of social network can provide further insight on their potential impact \cite{stella2018bots,brito2020complex,davis2020phase,pulido2020twitter,thelwall2020retweeting}, which calls for future approaches merging social networks, cognitive networks and emotional profiling.

Finally, understanding the impact of nuanced emotional appraisals would also benefit from investigating how these are related to behavioural and societal outcomes including the numbers of the contagion (e.g. hospitalisations, death rate, etc.) and compliance with physical distancing \cite{bedford2020covid}.

\section{Conclusions}
Given the massive attention devoted to the COVID-19 pandemic by social media, monitoring online discourse can offer an insightful thermometer of how individuals discussed and perceived the pandemic and the subsequent lockdown. Our MERCURIAL framework offered quantitative readings of the emotional profiles among Italian twitter users during early COVID-19 diffusion.  The detected emotional signals of political and social denounce, the trust in local authorities, the fear and anger towards the health and economic repercussions, and the positive initiatives of fraternity, all outline a rich picture of emotional reactions from Italians. Importantly, the psychological interpretation of MERCURIAL's results identified early signals of mental health distress and antisocial behaviour, both linked to violence and relevant for explaining increments in domestic abuse. Future research will further explore and consolidate the behavioural implications of online cognitive and emotional profiles, relying on the promising significance of our current results. Our cognitive network science approach offers decision-makers the prospect of being able to successfully detect global issues and design timely, data-informed policies. Especially under a crisis, when time constraints and pressure prevent even the richest and most organised governments from fully understanding the implications of their choices, an ethical and accurate monitoring of online discourses and emotional profiles constitutes an incredibly powerful support for facing global threats.

\section{Acknowledgements}
M.S. acknowledges Daniele Quercia, Nicola Perra and Andrea Baronchelli for stimulating discussion.

\section{Data Availability}
The IDs of the tweets analysed in this study are available on the Open Science Foundation repository: \url{https://osf.io/jy5kz/}.


\begin{thebibliography}{1}

\bibitem{zarocostas2020fight}
John Zarocostas.
\newblock How to fight an infodemic.
\newblock {\em The Lancet}, 395(10225):676, 2020.

\bibitem{cinelli2020covid}
Matteo Cinelli, Walter Quattrociocchi, Alessandro Galeazzi, Carlo~Michele
  Valensise, Emanuele Brugnoli, Ana~Lucia Schmidt, Paola Zola, Fabiana Zollo,
  and Antonio Scala.
\newblock The covid-19 social media infodemic.
\newblock {\em arXiv preprint arXiv:2003.05004}, 2020.

\bibitem{gallotti2020assessing}
Riccardo Gallotti, Francesco Valle, Nicola Castaldo, Pierluigi Sacco, and
  Manlio De~Domenico.
\newblock Assessing the risks of "infodemics" in response to {COVID}-19
  epidemics.
\newblock {\em arXiv preprint arXiv:2004.03997}, 2020.

\bibitem{pulido2020twitter}
Cristina~M Pulido, Beatriz Villarejo-Carballido, Gisela Redondo-Sama, and Aitor
  Gomez.
\newblock {COVID}-19 infodemic: More retweets for science-based information on
  coronavirus than for false information.
\newblock {\em International Sociology}, 0(0):0268580920914755, 2020.

\bibitem{Wang2020psychological}
C.~Wang, R.~Pan, X.~Wan, Y.~Tan, L.~Xu, C.S. Ho, and R.C Ho.
\newblock Immediate psychological responses and associated factors during the
  initial stage of the 2019 coronavirus disease ({COVID}-19) epidemic among the
  general population in {C}hina.
\newblock {\em International Journal of Environmental Research and Public
  Health}, 17:1729, 2020.

\bibitem{WHO2020mentalhealth}
{World Health Organization}.
\newblock Mental health during {COVID}-19 outbreak.
\newblock {\em Technical Report}, 2020.

\bibitem{Zhu2020mentalhealth}
S.~Zhu, Y.~Wu, C.Y Zhu, W.C. Hong, Z.X Yu, Z.K Chen, and Y.G. Wang.
\newblock The immediate mental health impacts of the {COVID}-19 pandemic among
  people with or without quarantine managements.
\newblock {\em Brain, Behavior, and Immunity}, IN PRESS, 2020.

\bibitem{Wang2020ptsd}
C.~Wang, R.~Pan, X.~Wan, Y.~Tan, L.~Xu, R.S. Mclntyre, F.N. Choo, B.~Tran,
  R.~Ho, V.K. Sharma, and C.~Ho.
\newblock A longitudinal study on the mental health of general population
  during the {COVID}-19 epidemic in {C}hina.
\newblock {\em Brain, Behavior, and Immunity}, IN PRESS, 2020.

\bibitem{ferrara2015quantifying}
Emilio Ferrara and Zeyao Yang.
\newblock Quantifying the effect of sentiment on information diffusion in
  social media.
\newblock {\em PeerJ Computer Science}, 1:e26, 2015.

\bibitem{davis2020phase}
Jessica~T Davis, Nicola Perra, Qian Zhang, Yamir Moreno, and Alessandro
  Vespignani.
\newblock Phase transitions in information spreading on structured populations.
\newblock {\em Nature Physics}, pages 1--7, 2020.

\bibitem{ciulla2012beating}
Fabio Ciulla, Delia Mocanu, Andrea Baronchelli, Bruno Gon{\c{c}}alves, Nicola
  Perra, and Alessandro Vespignani.
\newblock Beating the news using social media: the case study of american idol.
\newblock {\em EPJ Data Science}, 1(1):8, 2012.

\bibitem{stella2018bots}
Massimo Stella, Emilio Ferrara, and Manlio De~Domenico.
\newblock Bots increase exposure to negative and inflammatory content in online
  social systems.
\newblock {\em Proceedings of the National Academy of Sciences},
  115(49):12435--12440, 2018.

\bibitem{bail2018exposure}
Christopher~A Bail, Lisa~P Argyle, Taylor~W Brown, John~P Bumpus, Haohan Chen,
  MB~Fallin Hunzaker, Jaemin Lee, Marcus Mann, Friedolin Merhout, and Alexander
  Volfovsky.
\newblock Exposure to opposing views on social media can increase political
  polarization.
\newblock {\em Proceedings of the National Academy of Sciences},
  115(37):9216--9221, 2018.

\bibitem{siew2019cognitive}
Cynthia~SQ Siew, Dirk~U Wulff, Nicole~M Beckage, and Yoed~N Kenett.
\newblock Cognitive network science: A review of research on cognition through
  the lens of network representations, processes, and dynamics.
\newblock {\em Complexity}, 2019.

\bibitem{stella2020text}
Massimo Stella.
\newblock Text-mining forma mentis networks reconstruct public perception of
  the stem gender gap in social media.
\newblock {\em arXiv preprint arXiv:2003.08835}, 2020.

\bibitem{amancio2015probing}
Diego~R Amancio.
\newblock Probing the topological properties of complex networks modeling short
  written texts.
\newblock {\em PloS one}, 10(2), 2015.

\bibitem{quercia2011our}
Daniele Quercia, Michal Kosinski, David Stillwell, and Jon Crowcroft.
\newblock Our twitter profiles, our selves: Predicting personality with
  twitter.
\newblock In {\em 2011 {IEEE} third international conference on privacy,
  security, risk and trust and 2011 {IEEE} third international conference on
  social computing}, pages 180--185. IEEE, 2011.

\bibitem{mohammad2010emotions}
Saif~M Mohammad and Peter~D Turney.
\newblock Emotions evoked by common words and phrases: Using mechanical turk to
  create an emotion lexicon.
\newblock In {\em Proceedings of the NAACL HLT 2010 workshop on computational
  approaches to analysis and generation of emotion in text}, pages 26--34.
  Association for Computational Linguistics, 2010.

\bibitem{mohammad2018semeval}
Saif Mohammad, Felipe Bravo-Marquez, Mohammad Salameh, and Svetlana
  Kiritchenko.
\newblock Semeval-2018 task 1: Affect in tweets.
\newblock In {\em Proceedings of the 12th International Workshop on Semantic
  Evaluation}, pages 1--17, 2018.

\bibitem{kleinberg2020measuring}
Bennett Kleinberg, Isabelle van~der Vegt, and Maximilian Mozes.
\newblock Measuring emotions in the {COVID}-19 real world worry dataset.
\newblock {\em arXiv preprint arXiv:2004.04225}, 2020.

\bibitem{brito2020complex}
Ana Caroline~Medeiros Brito, Filipi~Nascimento Silva, and Diego~Raphael
  Amancio.
\newblock A complex network approach to political analysis: Application to the
  {B}razilian {C}hamber of {D}eputies.
\newblock {\em Plos one}, 15(3):e0229928, 2020.

\bibitem{plutchik1991emotions}
Robert Plutchik.
\newblock {\em The Emotions}.
\newblock University Press of America, 1991.

\bibitem{ekman1994nature}
Paul~Ed Ekman and Richard~J Davidson.
\newblock {\em The nature of emotion: Fundamental questions.}
\newblock Oxford University Press, USA, 1994.

\bibitem{hatfield1993emotional}
Elaine Hatfield, John~T Cacioppo, and Richard~L Rapson.
\newblock Emotional contagion.
\newblock {\em Current Directions in Psychological Science}, 2(3):96--100, 1993.

\bibitem{barsade2002ripple}
Sigal~G Barsade.
\newblock The ripple effect: Emotional contagion and its influence on group
  behavior.
\newblock {\em Administrative Science Quarterly}, 47(4):644--675, 2002.

\bibitem{kramer2014experimental}
Adam~DI Kramer, Jamie~E Guillory, and Jeffrey~T Hancock.
\newblock Experimental evidence of massive-scale emotional contagion through
  social networks.
\newblock {\em Proceedings of the National Academy of Sciences},
  111(24):8788--8790, 2014.

\bibitem{frey2019rippling}
Seth Frey, Karsten Donnay, Dirk Helbing, Robert~W Sumner, and Maarten~W Bos.
\newblock The rippling dynamics of valenced messages in naturalistic youth
  chat.
\newblock {\em Behavior Research Methods}, 51(4):1737--1753, 2019.

\bibitem{jasper2011emotions}
James~M Jasper.
\newblock Emotions and social movements: Twenty years of theory and research.
\newblock {\em Annual Review of Sociology}, 37:285--303, 2011.

\bibitem{stella2020forma}
Massimo Stella, Sarah De~Nigris, Aleksandra Aloric, and Cynthia~SQ Siew.
\newblock Forma mentis networks quantify crucial differences in stem perception
  between students and experts.
\newblock {\em PloS one}, 14(10), 2019.

\bibitem{posner2005circumplex}
Jonathan Posner, James~A Russell, and Bradley~S Peterson.
\newblock The circumplex model of affect: An integrative approach to affective
  neuroscience, cognitive development, and psychopathology.
\newblock {\em Development and Psychopathology}, 17(3):715--734, 2005.

\bibitem{de2016large}
Simon De~Deyne, Yoed~N Kenett, David Anaki, Miriam Faust, and Daniel~J Navarro.
\newblock Large-scale network representations of semantics in the mental
  lexicon.
\newblock {\em Big data in Cognitive Science: From Methods to Insights}, pages
  174--202, 2016.

\bibitem{stella2019innovation}
Massimo Stella and Anna Zaytseva.
\newblock Forma mentis networks map how nursing and engineering students
  enhance their mindsets about innovation and health during professional
  growth.
\newblock {\em PeerJ Computer Science}, 6:e255, 2020.

\bibitem{mehler2020topic}
Alexander Mehler, Rudiger Gleim, Regina Gaitsch, Wahed Hemati, and Tolga Uslu.
\newblock From topic networks to distributed cognitive maps: Zipfian topic
  universes in the area of volunteered geographic information.
\newblock {\em Complexity}, 2020, 2020.

\bibitem{stella2018distance}
Massimo Stella and Manlio De~Domenico.
\newblock Distance entropy cartography characterises centrality in complex
  networks.
\newblock {\em Entropy}, 20(4):268, 2018.

\bibitem{chen2020covid}
Emily Chen, Kristina Lerman, and Emilio Ferrara.
\newblock Covid-19: The first public coronavirus twitter dataset.
\newblock {\em arXiv preprint arXiv:2003.07372}, 2020.

\bibitem{marinho2018labelled}
Vanessa~Queiroz Marinho, Graeme Hirst, and Diego~Raphael Amancio.
\newblock Labelled network subgraphs reveal stylistic subtleties in written
  texts.
\newblock {\em Journal of Complex Networks}, 6(4):620--638, 2018.

\bibitem{vankrunkelsven2018predicting}
Hendrik Vankrunkelsven, Steven Verheyen, Gert Storms, and Simon De~Deyne.
\newblock Predicting lexical norms: A comparison between a word association
  model and text-based word co-occurrence models.
\newblock {\em Journal of {C}ognition}, 1(1), 2018.

\bibitem{newman2018networks}
Mark Newman.
\newblock {\em Networks}.
\newblock Oxford university press, 2018.

\bibitem{stella2019modelling}
Massimo Stella.
\newblock Modelling early word acquisition through multiplex lexical networks
  and machine learning.
\newblock {\em Big Data and Cognitive Computing}, 3(1):10, 2019.

\bibitem{warriner2013norms}
Amy~Beth Warriner, Victor Kuperman, and Marc Brysbaert.
\newblock Norms of valence, arousal, and dominance for 13,915 {E}nglish lemmas.
\newblock {\em Behavior {R}esearch {M}ethods}, 45(4):1191--1207, 2013.

\bibitem{russell1980circumplex}
James~A Russell.
\newblock A circumplex model of affect.
\newblock {\em Journal of Personality and Social Psychology}, 39(6):1161, 1980.

\bibitem{unwin2002effects}
Margaret~M Unwin, Dianna~T Kenny, and Pamela~J Davis.
\newblock The effects of group singing on mood.
\newblock {\em Psychology of Music}, 30(2):175--185, 2002.

\bibitem{fonagy2018affect}
Peter Fonagy, Gyorgy Gergely, and Elliot~L Jurist.
\newblock {\em Affect regulation, mentalization and the development of the
  self}.
\newblock Routledge, 2018.

\bibitem{stella2020formab}
Massimo Stella.
\newblock Forma mentis networks reconstruct how italian high schoolers and
  international stem experts perceive teachers, students, scientists, and
  school.
\newblock {\em Education {S}ciences}, 10(1):17, 2020.

\bibitem{miller1998wordnet}
George~A Miller.
\newblock {\em Word{N}et: An electronic lexical database}.
\newblock MIT Press, USA, 1998.

\bibitem{castro2019multiplex}
Nichol Castro and Massimo Stella.
\newblock The multiplex structure of the mental lexicon influences picture
  naming in people with aphasia.
\newblock {\em Journal of {C}omplex {N}etworks}, 7(6):913--931, 2019.

\bibitem{brugnoli2019recursive}
Emanuele Brugnoli, Matteo Cinelli, Walter Quattrociocchi, and Antonio Scala.
\newblock Recursive patterns in online echo chambers.
\newblock {\em Scientific Reports}, 9(1):1--18, 2019.

\bibitem{krueger1998perception}
Joachim Krueger.
\newblock On the perception of social consensus.
\newblock In {\em Advances in Experimental Social Psychology}, volume~30, pages
  163--240. Elsevier, 1998.

\bibitem{pearce2015ice}
Eiluned Pearce, Jacques Launay, and Robin~IM Dunbar.
\newblock The ice-breaker effect: Singing mediates fast social bonding.
\newblock {\em Royal Society Open Science}, 2(10):150221, 2015.

\bibitem{christensen2019just}
Julia~F Christensen, Steven Di~Costa, Brianna Beck, and Patrick Haggard.
\newblock I just lost it! fear and anger reduce the sense of agency: a study
  using intentional binding.
\newblock {\em {Experimental Brain Research}}, 237(5):1205--1212, 2019.

\bibitem{dewall2011general}
C~Nathan DeWall, Craig~A Anderson, and Brad~J Bushman.
\newblock The general aggression model: Theoretical extensions to violence.
\newblock {\em Psychology of Violence}, 1(3):245, 2011.

\bibitem{polanyi2006contextual}
Livia Polanyi and Annie Zaenen.
\newblock Contextual valence shifters.
\newblock In {\em Computing attitude and affect in text: Theory and
  applications}. Springer, 2006.

\bibitem{vad-acl2018}
Saif~M. Mohammad.
\newblock Obtaining reliable human ratings of valence, arousal, and dominance
  for 20,000 {E}nglish words.
\newblock In {\em Proceedings of The Annual Conference of the Association for
  Computational Linguistics (ACL)}, Melbourne, Australia, 2018.

\bibitem{kuppens2017relation}
Peter Kuppens, Francis Tuerlinckx, Michelle Yik, Peter Koval, Joachim
  Coosemans, Kevin~J Zeng, and James~A Russell.
\newblock The relation between valence and arousal in subjective experience
  varies with personality and culture.
\newblock {\em Journal of {P}ersonality}, 85(4):530--542, 2017.

\bibitem{vergallito2019somatic}
Alessandra Vergallito, Marco~Alessandro Petilli, Luigi Cattaneo, and Marco
  Marelli.
\newblock Somatic and visceral effects of word valence, arousal and
  concreteness in a continuum lexical space.
\newblock {\em Scientific Reports}, 9(1):1--10, 2019.

\bibitem{thelwall2020retweeting}
Mike Thelwall and Saheeda Thelwall.
\newblock Retweeting for {COVID}-19: Consensus building, information sharing,
  dissent, and lockdown life.
\newblock {\em arXiv preprint arXiv:2004.02793}, 2020.

\bibitem{bedford2020covid}
Juliet Bedford, Delia Enria, Johan Giesecke, David~L Heymann, Chikwe Ihekweazu,
  Gary Kobinger, H~Clifford Lane, Ziad Memish, Myoung-don Oh, Anne Schuchat,
  et~al.
\newblock {COVID}-19: towards controlling of a pandemic.
\newblock {\em The Lancet}, 2020.

\end{thebibliography}

\end{document}